\documentclass[conference]{IEEEtran}

\usepackage{multirow}
\usepackage{subfigure}

\usepackage[pdftex]{graphicx}
\usepackage{amsmath}
\usepackage{todonotes}

\ifCLASSINFOpdf
 
\else

\fi
 
\begin{document}

\title{Programmable Memristive Threshold Logic Gate Array}

\author{\IEEEauthorblockN{Olga Krestinskaya}
\IEEEauthorblockA{Department of \\ Electrical and Computer Engineering\\
Nazarbayev University,
Astana\\Email: ok@ieee.org}
\and
\IEEEauthorblockN{Akshay Kumar Maan}
\IEEEauthorblockA{Queensland Microelectronic Facility \\Griffith University\\
Queensland 4111, Australia\\
Email: akshay.maan@gmail.com}
\and
 \IEEEauthorblockN{Alex Pappachen James}
 \IEEEauthorblockA{Department of  \\Electrical and Computer  Engineering\\
Nazarbayev University,
Astana\\
Email: apj@ieee.org}}

\maketitle

\begin{abstract}
This paper proposes the implementation of programmable threshold logic gate (TLG) crossbar array based on modified TLG cells for high speed processing and computation. The proposed TLG array operation does not depend on input signal and time pulses, comparing to the existing architectures. The circuit is implemented using TSMC $180nm$ CMOS technology. The on-chip area and power dissipation of the simulated $3\times 4$ TLG array is $1463 \mu m^2$ and $425 \mu W$, respectively.
\end{abstract}

\IEEEpeerreviewmaketitle

\section{Introduction}

The implementation of memristive TLG cell and TLG array is an open problem that attracted a lot of interest in recent years \cite{imply,maan2017survey,magic,array}.
The memristive TLG array is an alternative solution for FPGA based processing that may allow to speed up computation and ensure small on-chip area and low power consumption \cite{array}. Therefore, in this paper, we propose the implementation of flexible programmable TLG array for high speed processing, where the memristor state is independent on the time and input to TLG cell.

There are several works that show the implementation of memristive TLG \cite{6415280,6331426,magic,maan2017survey,7803980} and memristive TLG arrays architecture \cite{imply,array}.
However, most of the existing designs are time-dependent, where the outputs are obtained in a sequence of pulses \cite{imply,magic} or memristor state depends on the inputs and should be reprogrammed while processing \cite{7803980}. 
Depending on the memristor material, the time to update a resistance of a memristor can vary from micro-seconds to seconds, and high voltage pulses are required to update the memristor value in a short time period \cite{borghetti2010memristive}, which cause the increase in power consumption. Therefore, the minimization of the number of pulses required for memristive TLG operation is an important issue in memristive TLG design \cite{wang2018synthesizing}.
To reduce the power dissipation required for processing and to speed up the computation, this paper proposes the implementation of programmable TLG array, where the operation does not depend on time and inputs, comparing to the existing architectures \cite{imply,array}.

The contributions of this work are the following:
\begin{itemize}
\item We proposed a modified implementation of threshold logic gates shown in \cite{7803980}, which have proven to be effective for various applications \cite{8023844,dastanova2018bit}. The modified TLG cell in independent on time and inputs.
\item We implement the programming circuit to update the threshold logic cell and analyze the limitations of non-ideal single transistor switches and their effect on the performance of the cell.
\item We present the design of flexible programmable TLG array with the proposed TLG cells and programming circuits that can be used for large-scale complex multiple threshold logic operations.
\end{itemize}
 
\section{Modified threshold logic gates}

\subsection{Threshold logic gates cell}

\begin{figure*}
\centering
\includegraphics[width=180mm]{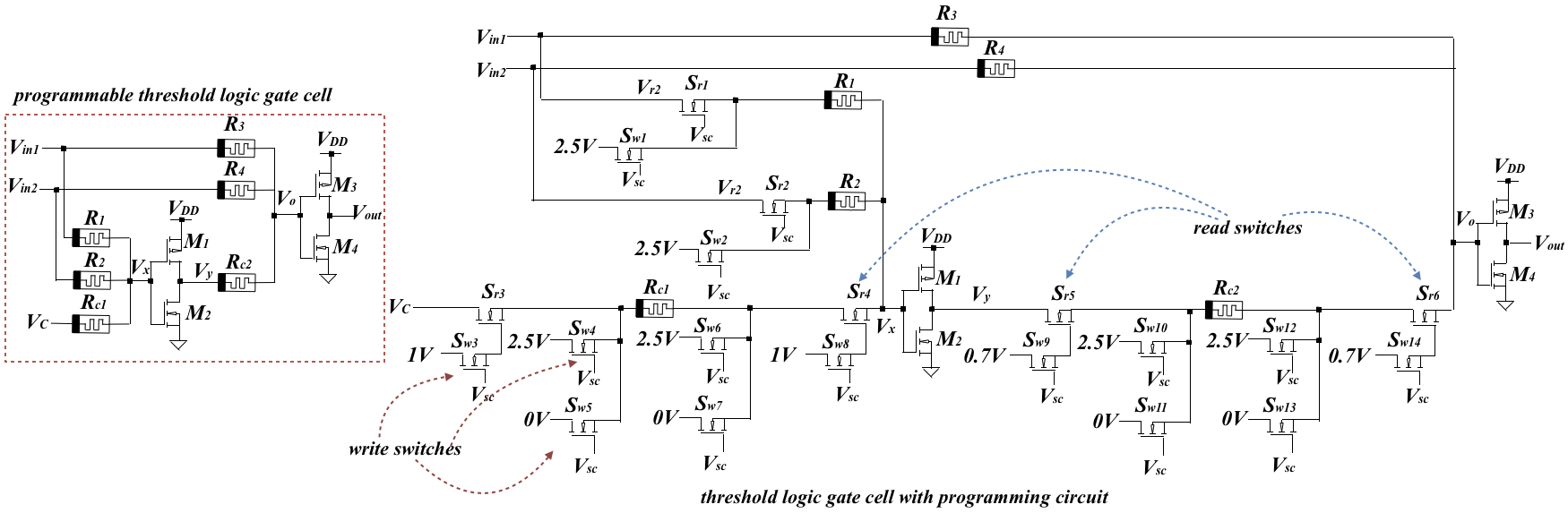}
\caption{Proposed circuit design for modified time independent programmable threshold logic gate cell (left size) inspired from \cite{7803980} and corresponding programming circuit (right side).}
\label{f1}
\end{figure*}

Figure \ref{f1} (left side) illustrates the implementation of TLG cell inspired from \cite{7803980}. The architecture is similar to the previously proposed TLG cell \cite{7803980}; however, the operating principle is modified. 
In previously proposed TLG cell, the memristor $R_{c2}$ and control signal $V_c$ is updated during the cell processing and change according to the input signal, which limits the processing speed,
especially when the time required to update the memristor is large. We proposed to implement unified time-independent TLG cell that can be easily integrated into TLG array, where memristors can be preprogrammed during the single step programming cycle, and the resistance and control voltage should not be changed based on the input. 
To implement time independent TLG, we modified the transistor parameter and control voltage signal. The threshold of the inverter $M3-M4$ is lowered by adjusting $W/L$ ratio of the transistors. New transistor parameters are the following:
$M1=0.72\mu/0.18\mu$,
$M2=0.36\mu/0.18\mu$,
$M3=0.72\mu/0.18\mu$,
$M4=20.36\mu/0.18\mu$. The memristors $R_{1}$, $R_{2}$, $R_{3}$ and $R_{4}$ are always programmed to high resistance $R_{off}$. The parameters of control memristors  $R_{c1}$ and $R_{c2}$ that control the operation of TLG  are shown in Table \ref{t0}. The memristor resistances used in this paper are $R_{on}=3k\Omega$
$R_{off}=60k\Omega$.

\begin{table}[h]
\centering
\caption{Parameters of control memristors $R_{c1}$ and $R_{c2}$ and control signal $V_c$ for modified TLG cell}
\label{t0}
\begin{tabular}{|l|l|l|l|}
\hline
\textbf{Configuration} & \textbf{$R_{c1}$} & \textbf{$R_{c1}$} & \textbf{$V_c$} \\ \hline
NAND                   & $R_{on}$ (low)    & $R_{off}$ (high)  & 0.8 V          \\ \hline
NOR                    & $R_{off}$ (high)  & $R_{off}$ (high)  & 0.8 V          \\ \hline
XNOR                   & $R_{off}$ (high)  & $R_{on}$ (low)    & 0.8 V          \\ \hline
\end{tabular}
\end{table}

\subsection{Programming circuit}
The implementation of the programming circuit for the proposed TLG cell is shown in Fig. \ref{f1} (right side).
The switches for read and write operation are implemented as single NMOS transistor. 
We use single transistor switch to keep on-chip area of the programming circuit small. Transistors $S_{r}$ are responsible for the read operation, and transistors $S_w$ are involved in the write operation and update of the value of the memristors.
 The $W/L$ ratio of the read and write transistors is
$M=10.36\mu/0.18\mu$. The width is increased to ensure linear performance.

This switch configuration is not ideal and the complete ON or OFF state is not possible. The non-ideal behavior of the switch effects the performance of the cell and voltage drop occurs.  
Therefore, we adjust $W/L$ ratio of the transistor
$M2=5.36\mu/0.18\mu$, which improves the performance of NMOS transistor for $180nm$ CMOS technology. However, this parameter should be adjusted based on the design technology.

To turn on the read transistors the voltage $1V$ for $S_{r1}$-$S_{r2}$ and $0.7V$ for $S_{r3}$-$S_{r4}$ is applied to the transistor gate. This voltage also causes the particular voltage drop, which is compensated by lowering the threshold of the inverter $M_1-M_2$, comparing to the cell without programming circuit. Both read and write operations are controlled by $V_{sc}$, which is either $2.5V$ or $0V$ for ON state and OFF state, respectively.
To update the memristors from high to low state, the $S_{w4}$ and $S_{w7}$ for $R_{c1}$ and $S_{w10}$ and $S_{w13}$ for $R_{c2}$ should be in ON state. For changing memristor state from low to high, the $S_{w5}$ and $S_{w6}$ for $R_{c1}$ and $S_{w11}$ and $S_{w12}$ for $R_{c2}$ should be in ON state. The update process of both memristors can be performed in parallel, as the switches in programming circuit allow to separate the control memristors.

\section{Threshold logic gate array architecture}

\subsection{Cell arrangement in threshold logic gate crossbar array}

\begin{figure}
\centering
\includegraphics[width=85mm]{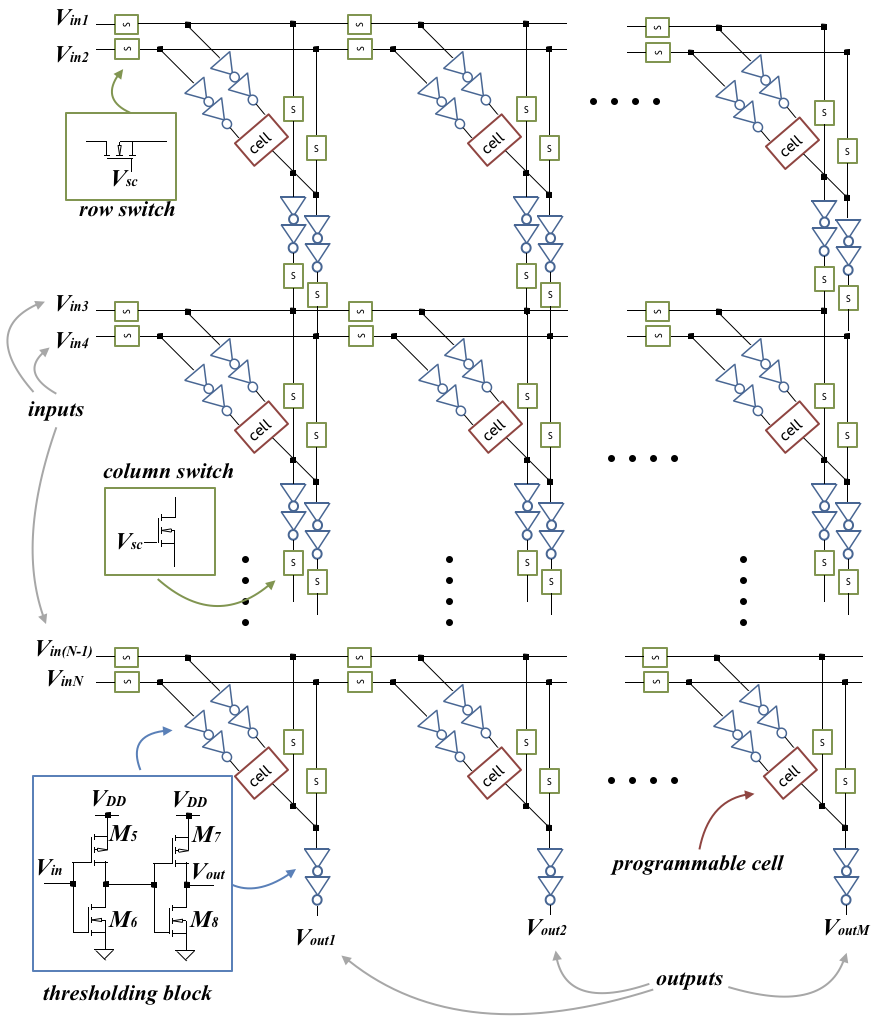}
\caption{Proposed TLG crossbar array configuration.}
\label{f2}
\end{figure}

The proposed TLG array architecture is shown in Fig. \ref{f2}. Each cell can be programmed to perform XNOR, XOR or NAND operation. The switching between rows and columns is performed by single transistor switches with $W/L=10.36\mu/0.18\mu$. Each row consists of two lines corresponding to a particular pair of inputs to the gates, which should be processed together. The inputs can be either connected or disconnected depending on the configuration. Each row corresponds to the particular inputs, this approach ensures the flexibility of the array programming, when the number of operations performed with the same pair of inputs is large. The outputs of the TLG cells are connected to two columns; this allows increase the number of possible configurations in the connection of neighboring cells. The number of lines in each row and column may be increase, if more complex functionality is required or the high array density should be achieved.

As in non-ideal switches the voltage drop may effect the performance and outputs of the TLG cells, we introduce the thresholding block consisting of two inverters with the following parameters: $M5=0.72\mu/0.18\mu$,
$M6=20.36\mu/0.18\mu$,
$M7=0.72\mu/0.18\mu$,
and $M8=0.36\mu/0.18\mu$. The threshold of this circuit is about 0.4 V. This reduces the effect of voltage drop across the switches to the array performance.  
The array is flexible in programming, and switches allow the routing of the TLG cell outputs in different directions.



\subsection{Cell modification to improve TLG array scalability}

\begin{figure}
\centering
\includegraphics[width=85mm]{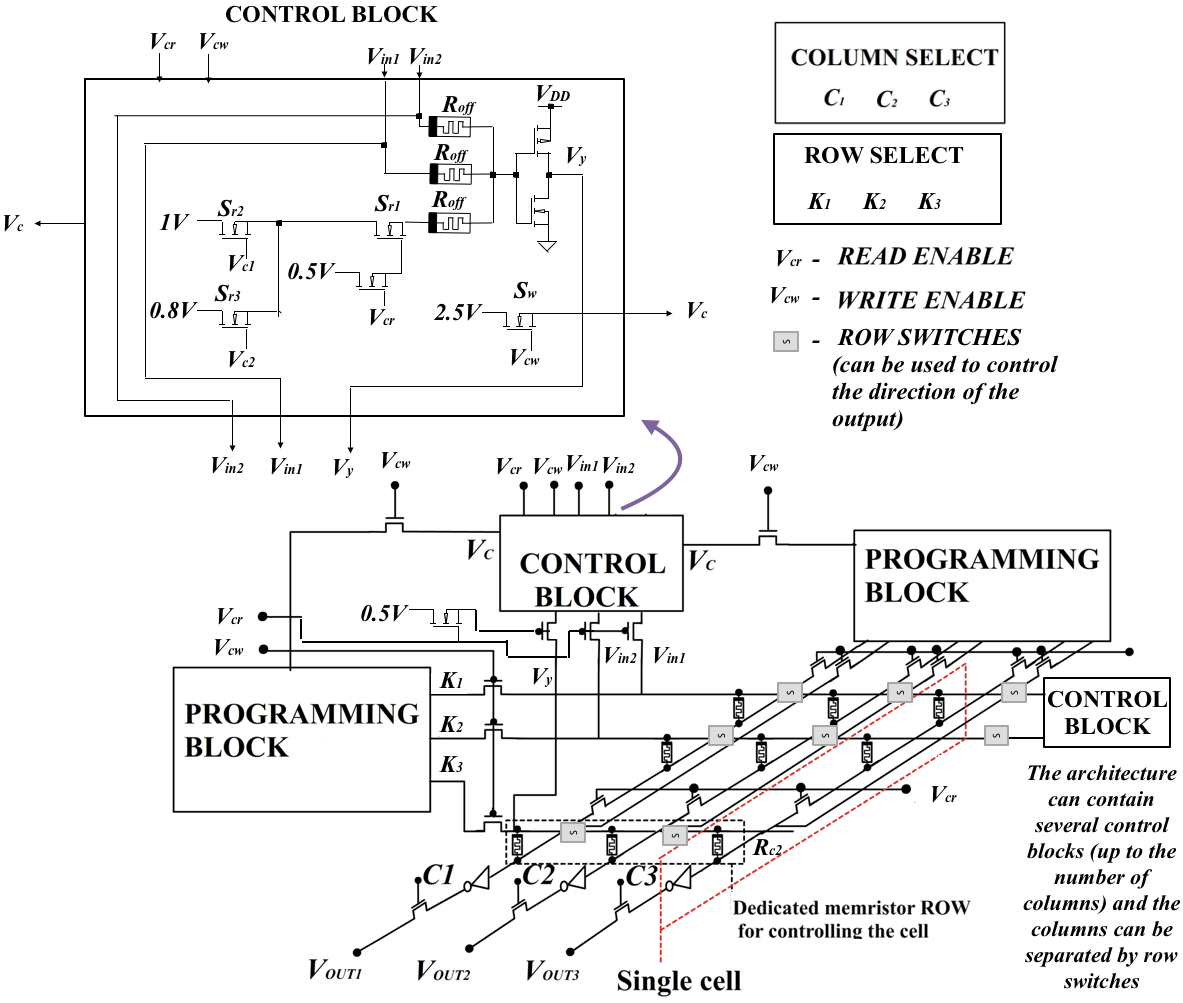}
\caption{Modification of the cell for programming circuit reduction and improvement of array arrangement.}
\label{aF}
\end{figure}

To improve ensure scalability of TLG array and reduce on-chip area of the programming circuit, the voltage control approach with single programmable memristor $R_{c2}$ and changing $V_{c}$ can be used. The truth table for this approach is shown in Table \ref{t1}. The memristors $R_1$, $R_2$ and $R_{c1}$ and inverter $M_1-M_2$ (Fig. \ref{f1}) can be separated from the architecture and switches corresponding to the programming of $R_{c1}$ can be removed. The modified cell architecture with the control of all lines is shown in Fig. \ref{aF}.
The programming of $R_{c2}$ can be performed by row and column transistors. 

The architecture is also appropriate for modular approach, when the TLG array`s cells are arranged into smaller arrays. This approach is useful when the processed data has a large number of various combinations of inputs that should be processed separately. In addition, the other possibility to use TLG cell illustrated in Fig. \ref{aF} is to preprogram the cells and control the functionality of the cell selecting particular rows and columns.

\begin{table}[h]
\centering
\caption{Parameters of control memristors $R_{c1}$ and $R_{c2}$ and control signal $V_c$ for TLG cell with a single programmable memristor and reduced programming circuit.}
\label{t1}
\begin{tabular}{|l|l|l|l|}
\hline
\textbf{Configuration} & \textbf{$R_{c1}$} & \textbf{$R_{c1}$} & \textbf{$V_c$} \\ \hline
NAND                   & $R_{off}$ (high)  & $R_{off}$ (high)  & 1 V            \\ \hline
NOR                    & $R_{off}$ (high)  & $R_{off}$ (high)  & 0.8 V          \\ \hline
XNOR                   & $R_{off}$ (high)  & $R_{on}$ (low)    & 0.8 V          \\ \hline
\end{tabular}
\end{table}

\section{Simulation results and discussion}

\begin{figure}
\centering
\includegraphics[width=90mm]{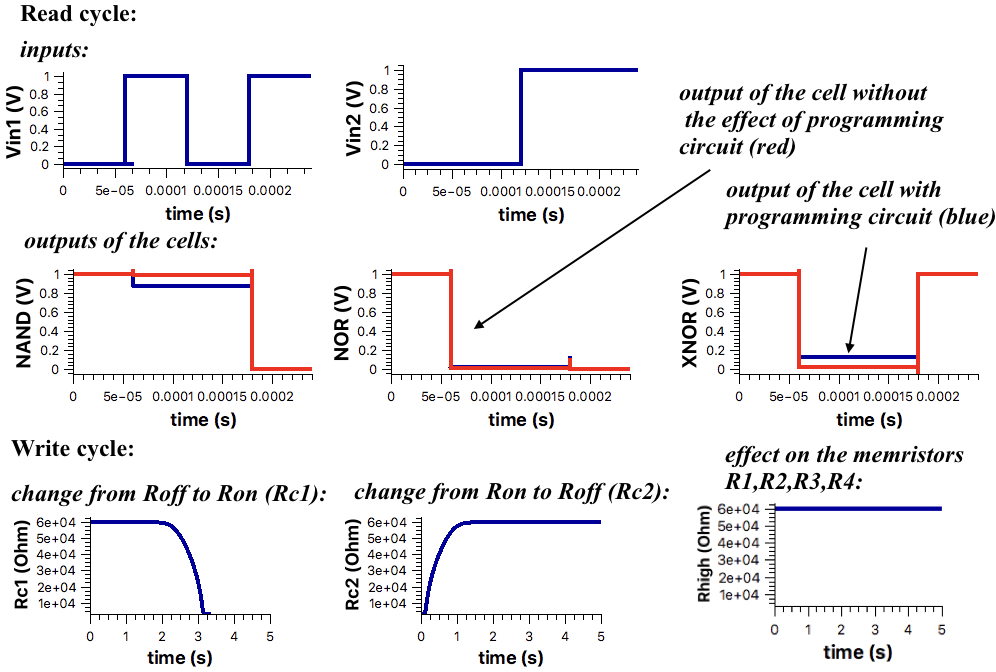}
\caption{Simulation of the proposed TLG cell with programming circuit.}
\label{f3}
\end{figure}

For this simulation, we used modified S memristor model shown in
\cite{biolek2018modeling} and TSMC $180nm$ CMOS technology. Also, we use underdrive inverters in the design with $V_{DD}=1V$ to ensure lower inverter threshold. The transient simulation of the proposed modified TLG cell for read and write operations are shown in Fig. \ref{f3}. In the read cycle, the outputs of three different cells are shown by red lines, and blue lines correspond to the outputs of the cell with programming circuit, which shows a small deviation from the ideal voltage in NAND and XNOR cells and can be improved using higher $R_{off}$ resistance or improving the performance of the read switches. Also, this deviation is easily removed by the thresholding circuit in the TLG crossbar array. In the writing cycle, the update process of memristors $R_{c1}$ for $R_{off}$ to $R_{on}$ and $R_{c2}$ from $R_{on}$ to $R_{off}$ is performed at the same time. It is shown that the update process of all the memristors for the proposed programming circuit can be done in parallel and does not effect the resistance of $R_1$, $R_2$, $R_3$ and $R_4$.

\begin{figure}
\centering
\includegraphics[width=90mm]{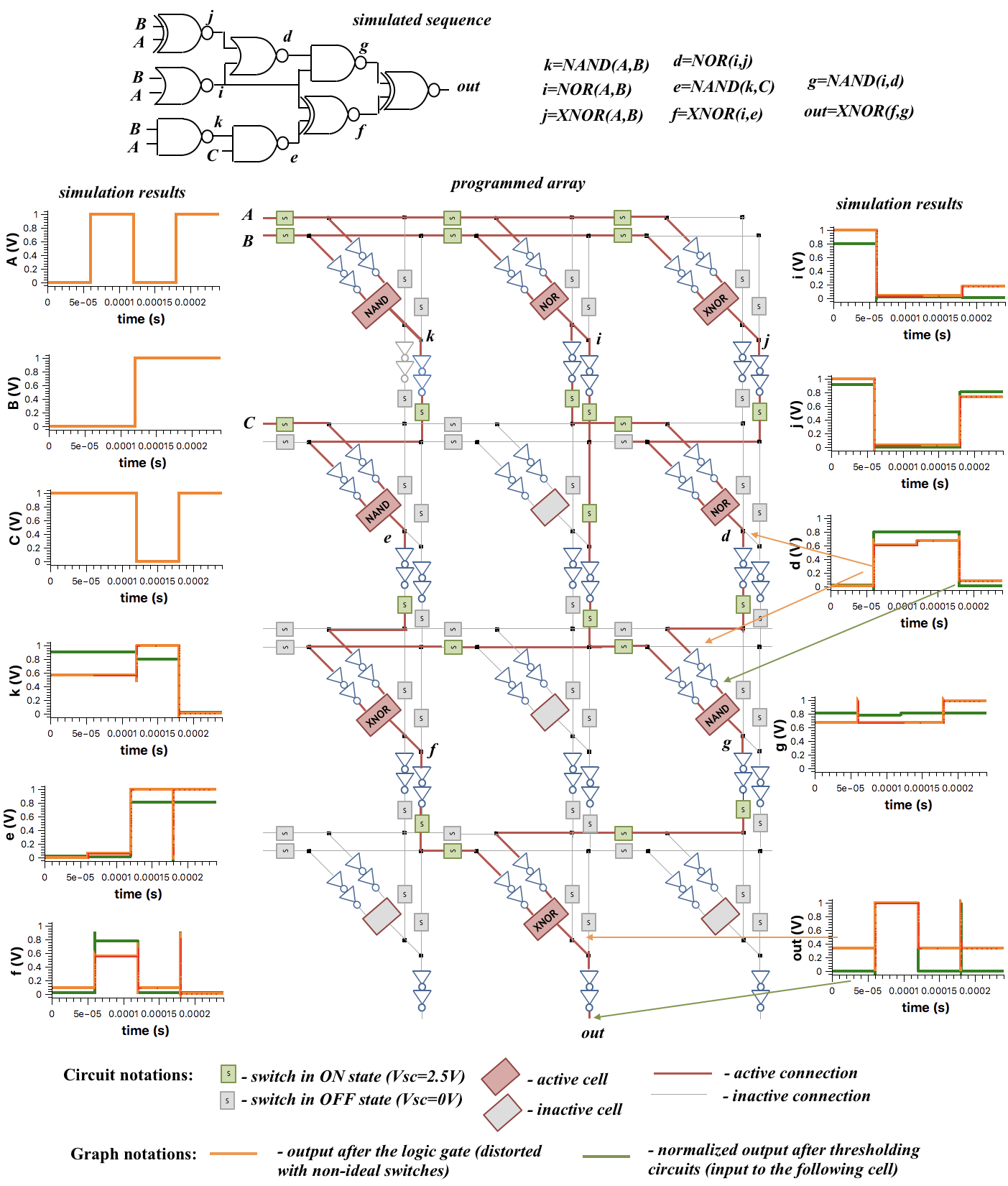}
\caption{Example of the performance of programmed TLG crossbar array.}
\label{f4}
\end{figure}

Figure \ref{f4} illustrate the example of the programming of $3\times 4$ TLG crossbar array programming and performance of the circuit. The exemplar sequence is shown in Fig. \ref{f4} (top). The output graphs show the outputs of the cell before (orange line) and after (green line) the thresholding block, which illustrates the thresholding block can compensate the voltage drop in the switches. The on-chip area and maximum power dissipation are shown in Table \ref{t2}. Comparing to the previous implementation \cite{7803980}, the on-chip area of the circuit TLG cell and programming circuit is improved.

\begin{table}[h]
\centering
\caption{Area and power calculation for the proposed TLG array}
\label{t2}
\begin{tabular}{|l|l|l|}
\hline
\textbf{Configuration}                                                           & \textbf{On-chip area}                                                                                                                   & \textbf{\begin{tabular}[c]{@{}l@{}}Maximum power \\ consumption\end{tabular}}          \\ \hline
Modified TLG cell                                                                & $7.863 \mu m^2$                                                                                                                         & \begin{tabular}[c]{@{}l@{}}NOR:$ 21.4 \mu W$, \\ NAND: $21.4 \mu W$, \\ XNOR:$30.8 \mu W$\end{tabular} \\ \hline

\begin{tabular}[c]{@{}l@{}}Proposed cell with\\ programming circuit\end{tabular} & $69.4662  \mu m^2$                                                                                                                                 & \begin{tabular}[c]{@{}l@{}}NOR:$20.56 \mu W$, \\ NAND: $20.48 \mu W$, \\ XNOR: $43.44 \mu W$\end{tabular}      \\ \hline

\begin{tabular}[c]{@{}l@{}}TLG cell with reduced\\ programming circuit\end{tabular} & $28.4281  \mu m^2$                                                                                                                                 & \begin{tabular}[c]{@{}l@{}}NOR:$15.72 \mu W$, \\ NAND: $10.42 \mu W$, \\ XNOR: $28.6 \mu W$\end{tabular}      \\ \hline

Simulated $3\times4$ array                                                        & \begin{tabular}[c]{@{}l@{}}thresholding block: \\ $7.9776 \mu m^2$\\ switch: $3.7296 \mu m^2$\\ array: $1462.6728 \mu m^2$\end{tabular} & $425.36 \mu W$                                                                         \\ \hline
\end{tabular}
\end{table}





Comparing to the existing threshold logic crossbar array architectures \cite{imply,array}, the advantage of the proposed architecture is time independent performance and absence of the resistors in the TLG cell. The crossbar array is programmed in a single cycle, and resistance of the memristors should not be change during the processing. The proposed array is adaptable and highly flexible, less complex and have smaller on-chip area, comparing to the traditional digital and FPGA-based solutions. The future work will include the improvement of the switch performance, investigation of the possibility to use memristive switch \cite{zhang2015novel}, development of the programming algorithm and analysis of the limitations of the proposed TLG array for large scale simulations. Also, the trade-off between power consumption, on-chip area and programming time will be investigated.

\section{Conclusion}

This work proposes the implementation of memristive programmable TLG crossbar array with modified TLG cell for high speed processing. The TLG cell was modified to be independent on the input and avoid changing the resistance of preprogrammed control memristor during the processing, which allows to speed up the operation. The proposed crossbar arrangement of TLG cells ensures flexibility in the programming. The non-ideal behavior of the devices has been considered. The on-chip area and power dissipation of the simulated $3\times 4$ TLG array is $1463 \mu m^2$ and $425 \mu W$, respectively. The future work includes the investigation of the array scalability, improvement of the performance in term of on-chip area and power dissipation and development of generalized crossbar programming algorithm.

\bibliographystyle{IEEEtran}
 \bibliography{ref}

\end{document}